\documentclass[a4paper,11pt]{article}
\usepackage{pos}
\usepackage{tabularx}
\usepackage{setspace}

\title{Techniques for Continuous Wave Identification and Filtering in the Askaryan Radio Array}
\ShortTitle{CW Identification \& Filtering in ARA}

\author*[1]{Mohammad Ful Hossain Seikh}
\author[1]{Dave Besson}
\author[2]{Pawan Giri} 
\onbehalf{for the ARA Collaboration \\{\normalsize \normalfont(a complete list of authors can be found at the end of the proceedings)}\\}

\affiliation[1]{Department of Physics \& Astronomy, University of Kansas, Lawrence, KS, USA}
\affiliation[2]{Department of Physics \& Astronomy, University of Nebraska--Lincoln, NE, USA}

\emailAdd{ $^*$fulhossain@ku.edu}
\emailAdd{zedlam@ku.edu}
\emailAdd{pgiri4@huskers.unl.edu}

\abstract{The Askaryan Radio Array (ARA), located near the geographical South Pole, is among the first experiments at the South Pole designed to detect ultra-high energy neutrinos through the Askaryan effect. When such neutrinos interact within dense media such as ice, they initiate particle cascades that, as they evolve, generate coherent radio pulses. Operating in the 150–850 MHz frequency band, ARA is deployed 80–200 meters deep in Antarctic ice, where the radio frequency background is exceptionally low. Despite the low background, experiments such as ARA must still account for continuous wave (CW) signals, which can originate from anthropogenic sources, instrumental noise, and other environmental factors. These CW signals can potentially obscure the faint neutrino-induced radio pulses, complicating data analysis and event identification. Over the years, ARA has developed and refined a number of techniques for CW filtering and identification, including spectral analysis, notch filtering, and phase-variance methods. These approaches exploit the unique characteristics of CW signals, such as their narrowband nature and temporal persistence, to effectively separate CW contamination from genuine impulsive events. We review the main CW identification and filtering techniques developed within the ARA collaboration and present recent improvements in their adaptive, multi-stage filtering pipelines. These advances have led to faster processing, easier operation, and more accurate CW identification and suppression, improving the consistency and quality of data analysis. The efficacy of these methods is demonstrated through CW identification and filtering for all ARA stations, showcasing their critical role in reducing event misclassification and improving the experiment's overall performance. By refining these techniques, this work not only improves the sensitivity and data analysis performance of ARA, but also underscores the importance of robust CW identification and filtering for current and future neutrino radio detection experiments.\\

{\bfseries Presenter:}
Marco Muzio$^{3}$\\
{$^{3}$ \itshape 
Department of Physics, University of Wisconsin-Madison, WI, USA
}

}

\ConferenceLogo{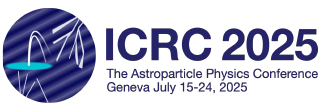}

\FullConference{39th International Cosmic Ray Conference (ICRC2025)\\
 15–24 July 2025\\
Geneva, Switzerland\\}

\ConferenceLogo{PoS_ICRC2025_logo.pdf}

\FullConference{39th International Cosmic Ray Conference (ICRC2025)\\
 15–24 July 2025\\
Geneva, Switzerland\\}

\begin{document}
\maketitle


\section{Introduction}
The search for ultra-high energy (UHE) neutrinos, with energies above $10^{15}$~eV, is one of the most compelling frontiers in astroparticle physics. UHE neutrinos are unique cosmic messengers: unlike photons or charged cosmic rays, they can traverse cosmic distances with minimal attenuation, carrying information from the most extreme environments in the Universe~\cite{Ackermann:2022, Vitagliano:2020}. Their detection would open a new window to both the origins of cosmic rays and the mechanisms of high-energy particle acceleration.

The Askaryan Radio Array (ARA) is one of the pioneering experiments designed to detect UHE neutrinos by exploiting the Askaryan effect, the coherent radio emission produced when UHE neutrinos interact in dense media such as polar ice~\cite{Askaryan:1965, Barwick:2023}. Deployed near the geographic South Pole, ARA comprises five autonomous stations embedded at depths of 80-200~m within the 2.8 km thick Antarctic ice sheet, operating over a radio frequency (RF) bandwidth of 150--850~MHz~\cite{Seikh:2024_EPJST, Allison:2012_AP}.

Maximizing sensitivity to rare Askaryan signals in ARA data requires effective suppression of backgrounds, including thermal noise and continuous wave (CW) interference originating from anthropogenic and environmental sources. While galactic noise is largely negligible above 150 MHz at the South Pole, persistent CW signals can obscure impulsive neutrino signals by raising the noise floor or contaminating specific frequency channels, thereby complicating event identification and potentially reducing experimental sensitivity~\cite{Mikhailova:2024, Meures:2016}.

This article presents a comprehensive overview of CW identification and filtering strategies developed and refined within the ARA Collaboration. We review the physical principles behind UHE neutrino detection, detail the sources and characteristics of radio backgrounds in the Antarctic environment, and summarize the state-of-the-art techniques, both classical and emerging, for robust CW mitigation. By advancing these methods, ARA not only enhances its discovery potential, but also establishes best practices for next-generation radio neutrino observatories.


\section{The ARA Detector}

ARA consists of five autonomous stations (A1--A5), each instrumented with deep in-ice antennas and calibration pulsers, deployed at depths of 80-200~m in Antarctic ice near the South Pole. The design, including antenna configuration, electronics, and calibration systems, has evolved across deployments to optimize sensitivity to impulsive radio signals and suppress backgrounds. A summary of the main instrument parameters is given in Table~\ref{tab:ARA_design}. ARA is continually improving its data acquisition and triggering systems, as well as developing new analysis techniques for robust identification and filtering of radio backgrounds, particularly CW interference.

\begin{table}[ht]
    \centering
    \caption{Key design parameters and instrumentation of the ARA stations~\cite{Seikh:2024_EPJST, Allison:2012_AP}.}
    \label{tab:ARA_design}
    \renewcommand{\arraystretch}{0.9}
    \begin{tabularx}{\linewidth}{|l|r|}
        \hline
        \textbf{Parameter} & \textbf{Value / Description}  \\
        \hline
        Number of stations & 5 (A1--A5), deployed 2012--2018  \\
        Station spacing & $\sim$1.6~km (average), grid layout \\
        Deployment depth & 80 (A1)--200 (A2--A5)~m (in-ice antennas) \\
        RX antennas per station & 16 total (deep): 8 VPol, 8 HPol\\
        TX (calpulser) antennas per station & 4 total (deep): 2 VPol, 2 HPol \\
        Surface antennas & 4 (A1--A3), none in A4, A5 \\
        Bandwidth & 150--850~MHz, bandpass + notch filters \\
        Antenna types & Birdcage (VPol), Quadslot (HPol) \\
        Amplification & $\sim$75~dB total gain (LNA + secondary) \\
        Digitizer & IRS2, 3.2~GS/s (Switched Capacitor Array) \\
        Triggering scheme & 3-of-8 (VPol or HPol); Phased array (A5) \\
        Clock/timing & GPS-synced rubidium oscillator \\
        Ice temperature (at depth) & $-40^\circ$C to $-50^\circ$C \\
        DAQ readout & Local storage; data transferred to IceCube Lab; sent North \\
        Trigger rate & RF: 5–10 Hz; Calpulser: 1 Hz; Software: 1 Hz (typical)\\
        \hline
    \end{tabularx}
\end{table}



\section{Radio Backgrounds in Ice-Based Neutrino Detectors}

Sensitive searches for UHE neutrino interactions require careful treatment of radio backgrounds, which can arise from a range of natural and anthropogenic sources within the detector's RF bandwidth. Understanding and mitigating these backgrounds is essential for maximizing sensitivity and avoiding false event identification~\cite{Mikhailova:2024, Seikh:2024_EPJST}. The principal categories of radio backgrounds observed by ARA and similar detectors are:
\begin{itemize}
    \item \textbf{Thermal noise:} Random voltage fluctuations arising from both the thermal motion of electrons in the antenna and receiver electronics, as well as thermal radiation from the surrounding ice and sky. This sets the fundamental noise floor and limits the minimum detectable signal amplitude. In-ice low antenna temperatures lead to exceptionally low thermal noise power densities~\cite{Allison:2012_AP}.
    \item \textbf{Galactic and atmospheric noise:} At the low end of the frequency band ($<$150~MHz), radio emission from the galactic plane and, less significantly, solar and atmospheric sources, can contribute to the total noise~\cite{Mikhailova:2024}. The effect is modulated by antenna pointing and time of day (sidereal variation).
    \item \textbf{Transient pulses:} Impulsive, short-duration signals from natural and man-made sources. Examples include solar radio bursts, triboelectric discharges during high wind events, and anthropogenic activities (e.g., machinery, overflights)~\cite{Clark:2018, Mikhailova:2024, Besson2023Triboelectric}. Some of these can mimic the expected neutrino-induced Askaryan pulses in both time and frequency domains.
    \item \textbf{Continuous waves (CW):} Narrowband, persistent signals mostly from anthropogenic sources such as radiosonde weather balloons, radio communications, satellites, and on-site electronics. These signals, if unfiltered, can dominate the frequency spectrum and obscure or bias the detection of genuine neutrino events~\cite{Seikh:2024_EPJST, Meures:2016}.
\end{itemize}

A detailed schematic of radio background classification can be found in Fig.~1 of Ref.~\cite{Mikhailova:2024}. Each type of background requires dedicated strategies for identification and filtering, as discussed in subsequent sections. Thermal noise sets the minimum threshold for signal detection and drives optimization of detector sensitivity. Galactic and atmospheric noise contribute mainly at the lowest frequencies and can exhibit daily modulation. Transient pulses and CWs represent the most significant sources of confusion for impulsive event searches, with CWs requiring particular attention due to their persistence and the diversity of their origins.


\subsection{CW Backgrounds: Sources and Characteristics}

These narrowband, long-duration signals can originate from a variety of anthropogenic and instrumental sources, and, if not mitigated, can obscure real signals, produce spurious detections, and reduce the reliability of spectral analyses~\cite{Seikh:2024_EPJST, Meures:2016}. CW signals are typically characterized by their well-defined frequencies and spectral persistence over time. In ARA, common signatures of CW contamination include sharp, narrow peaks in the frequency spectrum, sinusoidal or periodic structure in the time-domain waveforms, and, in most cases, phase coherence between channels. Some CWs are present nearly continuously, while others are transient, corresponding to operational schedules of nearby equipment or environmental factors. Table~\ref{tab:CW_sources_ARA} summarizes typical CW sources detected within the ARA bandwidth, their likely origins, and characterization. This table draws from South Pole communication frequency lists~\cite{SP_FreqList, SP_SXXIList}, ARA analyses~\cite{Seikh:2024_EPJST, Meures:2016}, and operational experience. Notably, certain frequencies (e.g., 124 MHz, 225 MHz, 143 MHz, 403 MHz) appear recurrently across stations and years, with some sources still unidentified.

\begin{table}[ht]
    \centering
    \caption{Common CW lines in the ARA frequency band, with known or suspected origins. Frequencies marked with an asterisk ($^\ast$) are routinely observed in ARA data.}
    \label{tab:CW_sources_ARA}
    \renewcommand{\arraystretch}{0.9}
    \begin{tabularx}{\linewidth}{|l|X|X|}
        \hline
        \textbf{Frequency (MHz)} & \textbf{Likely Source / Usage} & \textbf{Notes / Observations} \\
        \hline
        118.2, 121.5, 126.2$^\ast$ & Air traffic comms (VHF) & Intermittent; aircraft approaches \\
        100, 200, 300, 500 & Digitizer or clock harmonics & Instrumental, seen in many runs \\
        139.5, 142.6--143.8$^\ast$ & Field party and science comms & VHF radios, routine operations \\
        143.0--143.4 & Shuttle, Helo ops, MacOps & Repeater and ground comms \\
        225$^\ast$ & Unknown (possible instrumental) & Persistent in several stations~\cite{Clark:2018thesis} \\
        403$^\ast$ & Radiosonde weather balloons            & Strong, periodic (balloon launches) \\
        450.05--456.9$^\ast$ & South Pole LMR trunking/base  & Multiple strong lines, see \cite{SP_SXXIList} \\
        454.025 & IceCube project (radio) & Occasional strong lines \\
        650$^\ast$, 809$^\ast$ & Power line or station electronics  & Seen in A1 and A4\\
        156.8, 161.95 & Ship comms (Palmer Station) & Generally out of main ARA range  \\
        $>$800 & Other station/experimental sources  & Variable, occasional peaks \\
        \hline
    \end{tabularx}
\end{table}

The identification and characterization of CW sources are ongoing efforts in ARA. Certain CW sources are well understood and can be reliably excised with notch filters or subtraction methods; others, especially unidentified sources or those whose frequencies drift over time, require adaptive or more sophisticated mitigation strategies. Further details and a complete, time-resolved inventory of observed CW lines can be found in internal technical notes, instrument logs~\cite{SP_FreqList, SP_SXXIList}, and doctoral theses~\cite{Clark:2018}.


\section{CW Identification Methods}\label{sec:section4}

Accurate identification of CW signals in data is a fundamental step in mitigating narrowband radio interference in ARA data. Over the years, multiple techniques have been implemented within the collaboration, including frequency-domain statistical analyses, and phase coherence tests. Preliminary studies of machine learning approaches, including convolutional neural networks applied to spectral data, are ongoing within ARA to further enhance CW identification in future analyses. The following subsections summarize the main methods currently in use, focusing on their mathematical foundations and practical implementation in the ARA analysis framework.

\subsection{Testbed Method}

The Testbed method is a robust, data-driven approach to identifying persistent CW contamination. The procedure begins by separating recorded events into three categories based on trigger type: RF (any physics signal exceeding the trigger threshold), calpulser (periodic in-ice calibration pulses), and software triggers (forced 1 Hz environmental noise events). For each trigger type, the Fast Fourier Transform (FFT) is computed for every event and channel in a run. The magnitude spectra are then averaged across all events of each trigger class, yielding three ``baseline''spectra $\langle B(f) \rangle$ that characterize the typical frequency content in the run for each trigger type. For a given event, the observed spectrum $S(f)$ is compared to the corresponding $\langle B(f) \rangle$ (matching trigger type), typically through a frequency-by-frequency ratio or difference. Frequencies where the event spectrum significantly exceeds the baseline, defined via a threshold, e.g., $R(f) = S(f)/\langle B(f) \rangle > T$ for some threshold $T$, and ratio $R(f)$ are flagged as CW-contaminated and are recorded as likely CW lines for subsequent filtering (see Section~5). This threshold can be tuned empirically to balance false positives and sensitivity. Mathematically, the steps can be summarized as:
\begin{align}
    \langle B_{\rm trig}(f) \rangle = \frac{1}{N_{\rm trig}} \sum_{e=1}^{N_{\rm trig}} |{\rm FFT}_e^{\rm trig}(f)| ; ~~~
    R(f) = \frac{|{\rm FFT}_e^{\rm trig}(f)|}{\langle B_{\rm trig}(f) \rangle} ; ~~~
    \text{CW flag at } f : R(f) > T
\end{align}
where $N_{\rm trig}$ is the number of events for a given trigger, and $|{\rm FFT}_e^{\rm trig}(f)|$ is the spectrum of event $e$.

\subsection{Phase Variance Method}

\begin{figure}[ht]
    \centering
    \includegraphics[width=0.48\linewidth]{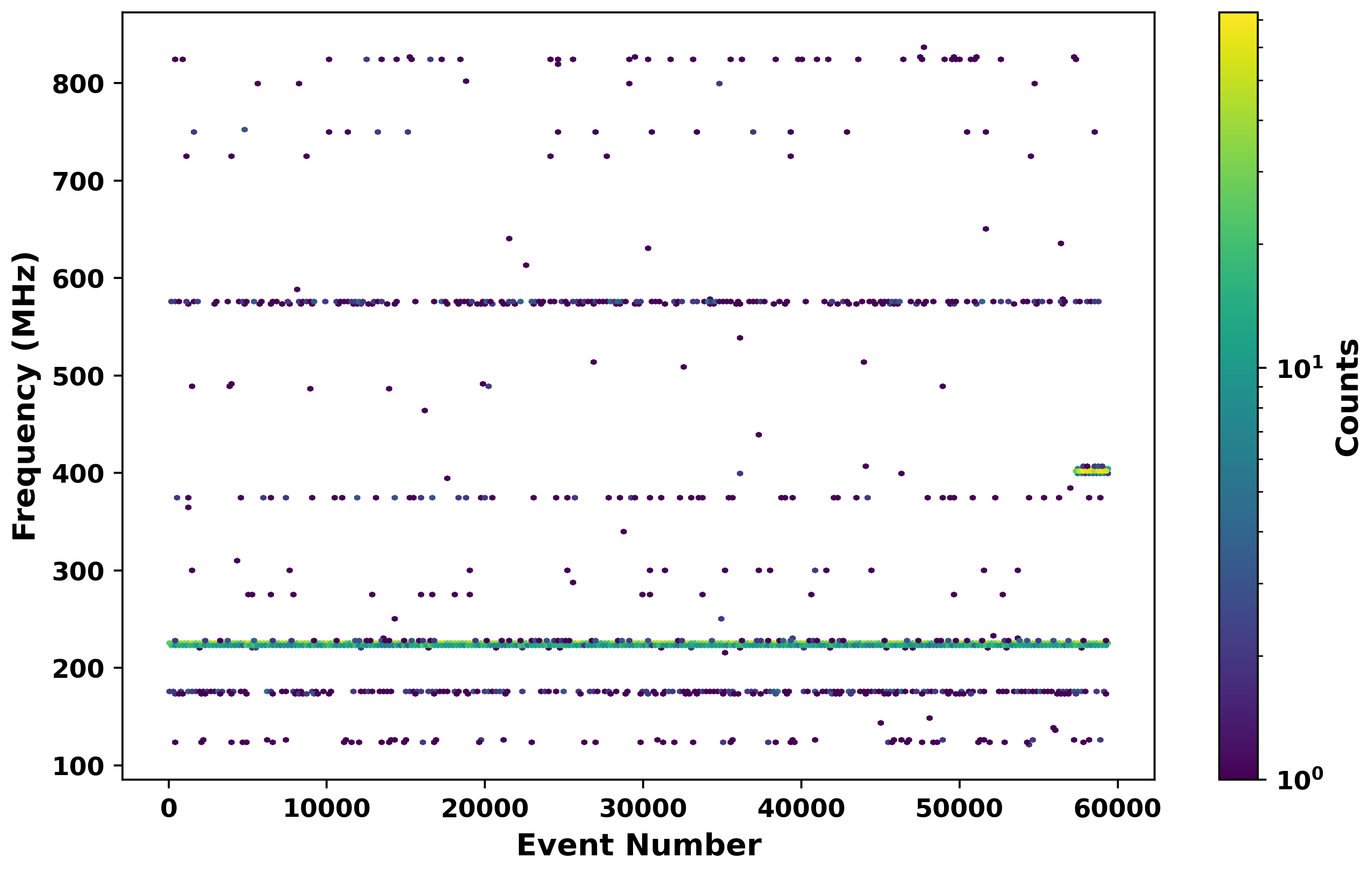}
    \hfill
    \includegraphics[width=0.48\linewidth]{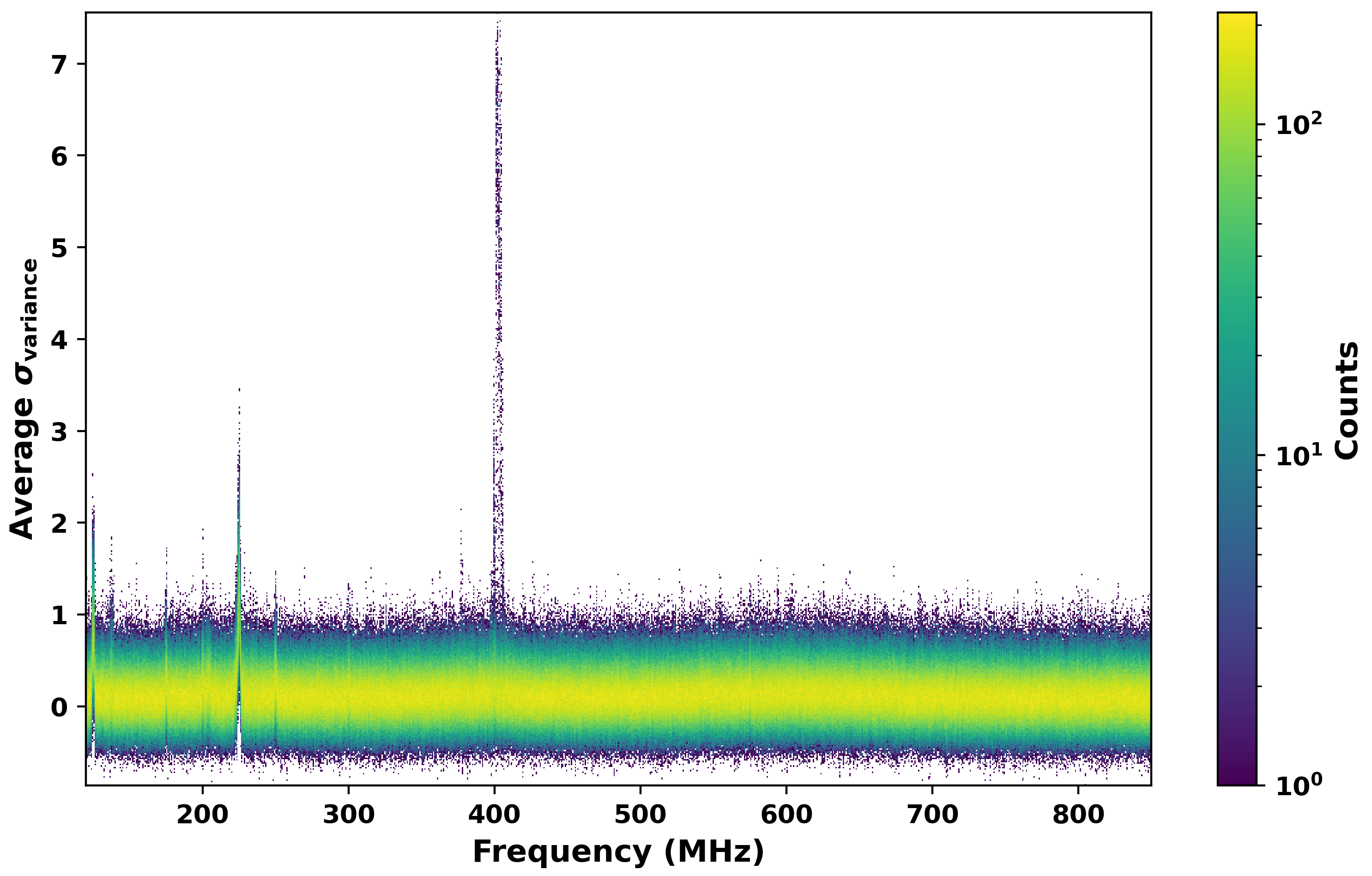}
    \caption{
        \textbf{Left:} Frequencies identified as CW-contaminated by the testbed algorithm for each event.
        \textbf{Right:} 2D histogram of average $\sigma_{\rm variance}$ as a function of frequency, showing power fluctuations and prominent narrowband CW features identified by phase variance method.
    }
    \label{fig:cwid}
\end{figure}

The Phase Variance Method provides a robust statistical approach for identifying CW contamination by leveraging the temporal coherence characteristic of CW signals. The central insight is that for a genuine impulsive signal or noise, the relative phase between channels (or channel pairs) will fluctuate randomly from event to event, while for a stable CW source, the phase difference will remain highly stable when evaluated over a sequence of consecutive events within a sliding window. Operationally, the method proceeds as follows: for each event and channel pair, the waveforms are calibrated, interpolated, dedispersed, and bandpass filtered. The FFT is computed, and the phase difference $\Delta\phi_{ij}(f,e)$ for frequency $f$, channel pair $(i,j)$, and event $e$ is extracted. A moving window of $N$ consecutive events (typically $N \sim 15$) is used to calculate the variance of  $\Delta\phi_{ij}(f,e)$ across the window for each channel pair and frequency. For each frequency $f$ and channel pair, the circular phase variance $V(f)$ is computed over a window of $N$ events,
\begin{align}
V(f) = 1 - \frac{1}{N} \left| \sum_{k=1}^N \exp(i\Delta\phi_{ij}(f,e + k)) \right|,
\end{align}
and the “sigma-variance” is
\begin{align}
\sigma_\mathrm{variance}(f) = \frac{\mathrm{median}(V(f)) - V(f)}{\sigma(f)},~~ \mathrm{with}~~ \sigma(f) = \frac{V_{95}(f) - \mathrm{median}(V(f))}{1.64}.
\end{align}
Here, $V_{95}(f)$ denotes the 95th percentile of the circular phase variance distribution at frequency $f$. The factor $1.64$ corresponds to the number of standard deviations between the median and the 95th percentile in a normal distribution. Frequencies with sigma-variance above threshold are identified as CW-contaminated. This process is performed for both forward and backward windows, and separately for vertical and horizontal polarization channel pairs. Compared to amplitude-based methods, the Phase Variance Method is particularly sensitive to weak or intermittent CWs, and complements other identification techniques for maximal detection efficiency. Refer to Section 4.8 of \cite{Clark:2018thesis} for further details. Fig.~\ref{fig:cwid} shows how Testbed and Phase Variance methods identify CW frequencies. 

\section{CW Filtering Techniques}

\subsection{SineSubtract Filters}

The SineSubtract filter is a time-domain technique designed to suppress narrowband CW interference in the ANITA dataset. Later it was re-tuned for the ARA data analysis. Candidate CW frequencies to filter can be drawn from the set of frequencies identified by the algorithms in Section~\ref{sec:section4}, or from a configurable list defined in an analysis. In both cases, a minimum power ratio (threshold), comparing the spectral power at each candidate frequency to the local spectral average, is required for a frequency to be flagged and subjected to filtering. This threshold is user-tunable and helps balance CW removal with signal preservation. For each selected frequency $f_{\rm CW}$, the voltage trace $V(t)$ is fit with a sinusoidal model,
\begin{align}
S(t) = a\sin(2\pi f_{\rm CW} t) + b\cos(2\pi f_{\rm CW} t),
\end{align}
where $a$ and $b$ are determined by linear least squares over the waveform or within a moving window. The fitted component $S(t)$ is then subtracted from $V(t)$, effectively removing the CW tone at that frequency. This process is repeated for all flagged frequencies, resulting in a filtered waveform with suppressed CW contamination and minimal distortion of underlying impulsive signals. The SineSubtract method is implemented following the FFTtools library~\cite{Deaconu:FFTtools}.

\subsection{Geometric Filtering}

Geometric filtering mitigates CW contamination by correcting both the amplitude and phase of contaminated frequency bins in the waveform’s spectrum. First, frequency bins identified as CW-contaminated, whether by phase variance, testbed methods, or visual inspections, are grouped into bands. For each contaminated frequency bin $f_k$, the spectral amplitude $A(f_k)$ is interpolated from neighboring clean bins, and the phase $\phi(f_k)$ is estimated as the geometric (rolling mean) phase of adjacent uncontaminated regions:
\begin{align}
A_{\rm clean}(f_k) = \mathrm{Interp}\left[A(f)\right]_{f \notin \text{CW}}, \quad
\phi_{\rm geom}(f_k) = \arg\left(\frac{1}{N}\sum_{j} A(f_j) e^{i\phi(f_j)}\right)
\end{align}
where $f_j$ are the frequencies in a window around $f_k$ not flagged as CW, and $N$ is the number of such bins. The contaminated spectral component is replaced by $A_{\rm clean}(f_k) e^{i\phi_{\rm geom}(f_k)}$, and the cleaned spectrum is inverse Fourier transformed to recover the time-domain waveform:
\begin{align}
V_{\rm cleaned}(t) = \mathrm{Re}\left[\mathcal{F}^{-1}\left\{A_{\rm clean}(f) e^{i\phi_{\rm geom}(f)}\right\}\right]
\end{align}
This approach allows robust CW suppression, retaining the spectral shape and impulsive signal content, while adaptively handling phase and amplitude anomalies in contaminated regions.


\section{Pipeline and Software}
All identification and filtering techniques described in this work, along with analysis scripts and documentation to enable reproducible results, are available in the Continuous Wave Identification and Suppression Ensemble for ARA (CWISE-ARA) software package, which is actively maintained and under continuous development.


\section{Results and Performance}
\begin{figure}[ht]
    \centering
    \includegraphics[width=0.9\linewidth]{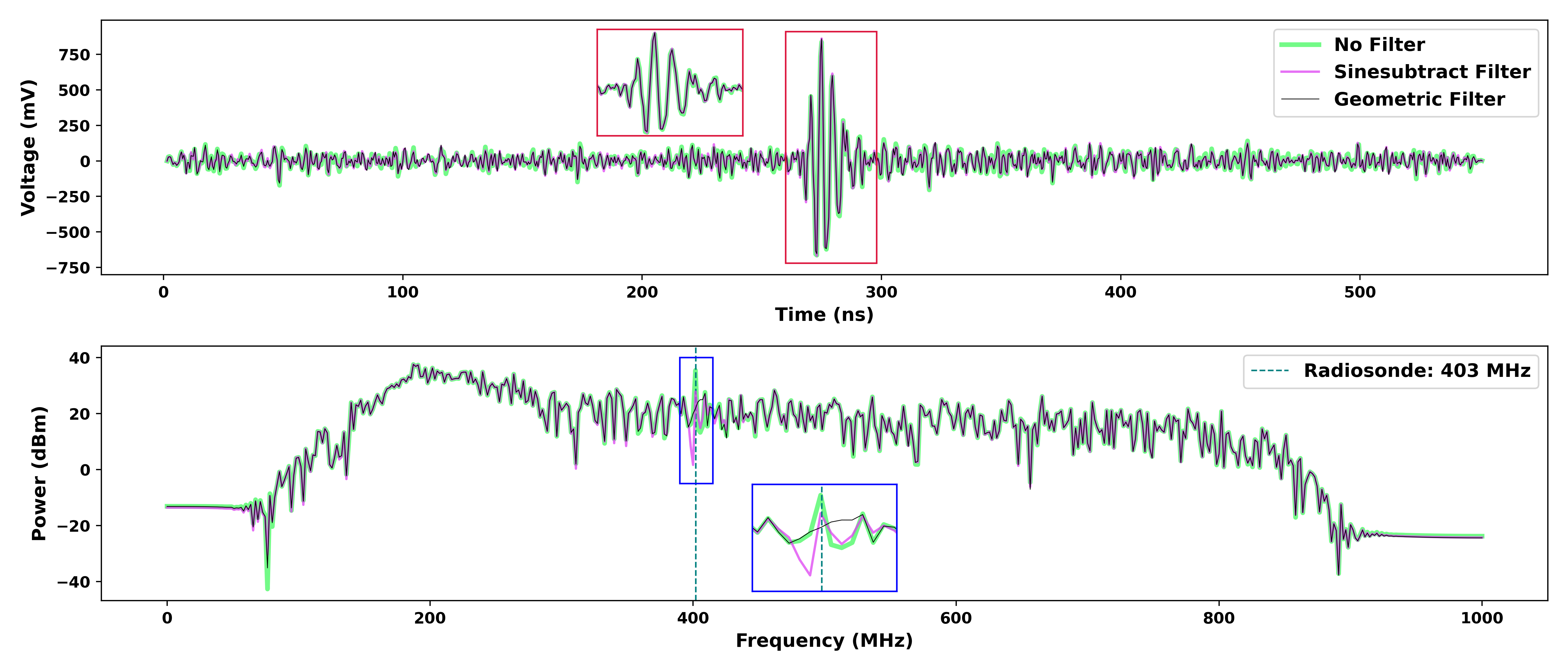}
    \caption{
        Comparison of filtering methods in a radiosonde contaminated calpulser trigger. {\bf Top}: Dedispersed waveforms with no filters, SineSubtract and geometric CW filters. {\bf Bottom}: Corresponding bandpassed filtered spectra. Zoomed-in boxes highlight the effects of CW filtering on the radiosonde lines, particularly visible in the spectra. 
    }
    \label{fig:cdfilter}
\end{figure}
The Testbed method is fast and effective for persistent CW lines, while Phase Variance is more sensitive to weak or intermittent signals but requires more computation. In practice, using both methods together provides the most comprehensive CW identification. Both the sine subtraction and geometric filtering techniques are effective at suppressing CW contamination. However, the geometric filter better preserves the original signal power, as shown in Fig.~\ref{fig:cdfilter}, particularly in the frequency-domain where it maintains the spectral amplitude closer to the unfiltered signal within the filtered bands. In contrast, the SineSubtract method tends to over-suppress and can introduce unnatural artifacts in the filtered frequency regions. In terms of computational performance, SineSubtract is faster since it directly removes identified tones, while geometric filtering is more computationally intensive due to the additional CW identification steps (testbed or phase variance) and interpolation procedures. Overall, all CW removal techniques can introduce spectral artifacts, but this can be partially mitigated by employing a diverse suite of removal techniques.


\section{Conclusion and Outlook}

Robust CW identification and filtering are essential for maintaining ARA’s sensitivity to rare neutrino signals. The geometric filtering method provides superior signal preservation, while sine subtraction offers speed and simplicity. Ongoing work will integrate adaptive and machine learning-based CW identification into the CWISE-ARA pipeline, with the goal of further improving efficiency, reliability, and automation for future data-taking and next-generation radio neutrino experiments.

Looking ahead, we plan to leverage CW identification for more than just event cleaning. Systematic CW studies, using advanced algorithms and machine learning, can reveal information about the detector environment, source localization, instrumental systematics, and site radio backgrounds. The CWISE-ARA framework will be extended to enable these analyses, making use of the identified CW features for environmental monitoring and potential calibration, in addition to their traditional role in signal filtering.

\begingroup
\setstretch{0.95}
\setlength{\bibsep}{1.0pt}
\bibliographystyle{ICRC}
{\small\providecommand{\href}[2]{#2}\begingroup\raggedright\endgroup
}
\endgroup

{\small 
\section*{Full Author List: ARA Collaboration (June 30, 2025)}

\noindent
N.~Alden\textsuperscript{1}, 
S.~Ali\textsuperscript{2}, 
P.~Allison\textsuperscript{3}, 
S.~Archambault\textsuperscript{4}, 
J.J.~Beatty\textsuperscript{3}, 
D.Z.~Besson\textsuperscript{2$\dagger$}, 
A.~Bishop\textsuperscript{5}, 
P.~Chen\textsuperscript{6}, 
Y.C.~Chen\textsuperscript{6}, 
Y.-C.~Chen\textsuperscript{6}, 
S.~Chiche\textsuperscript{7}, 
B.A.~Clark\textsuperscript{8}, 
A.~Connolly\textsuperscript{3}, 
K.~Couberly\textsuperscript{2}, 
L.~Cremonesi\textsuperscript{9}, 
A.~Cummings\textsuperscript{10,11,12}, 
P.~Dasgupta\textsuperscript{3}, 
R.~Debolt\textsuperscript{3}, 
S.~de~Kockere\textsuperscript{13}, 
K.D.~de~Vries\textsuperscript{13}, 
C.~Deaconu\textsuperscript{1}, 
M.A.~DuVernois\textsuperscript{5}, 
J.~Flaherty\textsuperscript{3}, 
E.~Friedman\textsuperscript{8}, 
R.~Gaior\textsuperscript{4}, 
P.~Giri\textsuperscript{14}, 
J.~Hanson\textsuperscript{15}, 
N.~Harty\textsuperscript{16}, 
K.D.~Hoffman\textsuperscript{8}, 
M.-H.~Huang\textsuperscript{6,17}, 
K.~Hughes\textsuperscript{3}, 
A.~Ishihara\textsuperscript{4}, 
A.~Karle\textsuperscript{5}, 
J.L.~Kelley\textsuperscript{5}, 
K.-C.~Kim\textsuperscript{8}, 
M.-C.~Kim\textsuperscript{4}, 
I.~Kravchenko\textsuperscript{14}, 
R.~Krebs\textsuperscript{10,11}, 
C.Y.~Kuo\textsuperscript{6}, 
K.~Kurusu\textsuperscript{4}, 
U.A.~Latif\textsuperscript{13}, 
C.H.~Liu\textsuperscript{14}, 
T.C.~Liu\textsuperscript{6,18}, 
W.~Luszczak\textsuperscript{3}, 
A.~Machtay\textsuperscript{3}, 
K.~Mase\textsuperscript{4}, 
M.S.~Muzio\textsuperscript{5,10,11,12$*$}, 
J.~Nam\textsuperscript{6}, 
R.J.~Nichol\textsuperscript{9}, 
A.~Novikov\textsuperscript{16}, 
A.~Nozdrina\textsuperscript{3}, 
E.~Oberla\textsuperscript{1}, 
C.W.~Pai\textsuperscript{6}, 
Y.~Pan\textsuperscript{16}, 
C.~Pfendner\textsuperscript{19}, 
N.~Punsuebsay\textsuperscript{16}, 
J.~Roth\textsuperscript{16}, 
A.~Salcedo-Gomez\textsuperscript{3}, 
D.~Seckel\textsuperscript{16}, 
M.F.H.~Seikh\textsuperscript{2$\dagger$}, 
Y.-S.~Shiao\textsuperscript{6,20}, 
S.C.~Su\textsuperscript{6}, 
S.~Toscano\textsuperscript{7}, 
J.~Torres\textsuperscript{3}, 
J.~Touart\textsuperscript{8}, 
N.~van~Eijndhoven\textsuperscript{13}, 
A.~Vieregg\textsuperscript{1}, 
M.~Vilarino~Fostier\textsuperscript{7}, 
M.-Z.~Wang\textsuperscript{6}, 
S.-H.~Wang\textsuperscript{6}, 
P.~Windischhofer\textsuperscript{1}, 
S.A.~Wissel\textsuperscript{10,11,12}, 
C.~Xie\textsuperscript{9}, 
S.~Yoshida\textsuperscript{4}, 
R.~Young\textsuperscript{2}
\\\\
$^\dagger$Corresponding Authors\\
$^*$Presenter
\\\\
\textsuperscript{1} Dept. of Physics, Enrico Fermi Institute, Kavli Institute for Cosmological Physics, University of Chicago, Chicago, IL 60637\\
\textsuperscript{2} Dept. of Physics and Astronomy, University of Kansas, Lawrence, KS 66045\\
\textsuperscript{3} Dept. of Physics, Center for Cosmology and AstroParticle Physics, The Ohio State University, Columbus, OH 43210\\
\textsuperscript{4} Dept. of Physics, Chiba University, Chiba, Japan\\
\textsuperscript{5} Dept. of Physics, University of Wisconsin-Madison, Madison,  WI 53706\\
\textsuperscript{6} Dept. of Physics, Grad. Inst. of Astrophys., Leung Center for Cosmology and Particle Astrophysics, National Taiwan University, Taipei, Taiwan\\
\textsuperscript{7} Universite Libre de Bruxelles, Science Faculty CP230, B-1050 Brussels, Belgium\\
\textsuperscript{8} Dept. of Physics, University of Maryland, College Park, MD 20742\\
\textsuperscript{9} Dept. of Physics and Astronomy, University College London, London, United Kingdom\\
\textsuperscript{10} Center for Multi-Messenger Astrophysics, Institute for Gravitation and the Cosmos, Pennsylvania State University, University Park, PA 16802\\
\textsuperscript{11} Dept. of Physics, Pennsylvania State University, University Park, PA 16802\\
\textsuperscript{12} Dept. of Astronomy and Astrophysics, Pennsylvania State University, University Park, PA 16802\\
\textsuperscript{13} Vrije Universiteit Brussel, Brussels, Belgium\\
\textsuperscript{14} Dept. of Physics and Astronomy, University of Nebraska, Lincoln, Nebraska 68588\\
\textsuperscript{15} Dept. Physics and Astronomy, Whittier College, Whittier, CA 90602\\
\textsuperscript{16} Dept. of Physics, University of Delaware, Newark, DE 19716\\
\textsuperscript{17} Dept. of Energy Engineering, National United University, Miaoli, Taiwan\\
\textsuperscript{18} Dept. of Applied Physics, National Pingtung University, Pingtung City, Pingtung County 900393, Taiwan\\
\textsuperscript{19} Dept. of Physics and Astronomy, Denison University, Granville, Ohio 43023\\
\textsuperscript{20} National Nano Device Laboratories, Hsinchu 300, Taiwan\\

\section*{Acknowledgements}

\noindent
The ARA Collaboration is grateful to support from the National Science Foundation through Award 2013134.
The ARA Collaboration
designed, constructed, and now operates the ARA detectors. We would like to thank IceCube, and specifically the winterovers for the support in operating the
detector. Data processing and calibration, Monte Carlo
simulations of the detector and of theoretical models
and data analyses were performed by a large number
of collaboration members, who also discussed and approved the scientific results presented here. We are
thankful to Antarctic Support Contractor staff, a Leidos unit 
for field support and enabling our work on the harshest continent. We thank the National Science Foundation (NSF) Office of Polar Programs and
Physics Division for funding support. We further thank
the Taiwan National Science Councils Vanguard Program NSC 92-2628-M-002-09 and the Belgian F.R.S.-
FNRS Grant 4.4508.01 and FWO. 
K. Hughes thanks the NSF for
support through the Graduate Research Fellowship Program Award DGE-1746045. A. Connolly thanks the NSF for
Award 1806923 and 2209588, and also acknowledges the Ohio Supercomputer Center. S. A. Wissel thanks the NSF for support through CAREER Award 2033500.
A. Vieregg thanks the Sloan Foundation and the Research Corporation for Science Advancement, the Research Computing Center and the Kavli Institute for Cosmological Physics at the University of Chicago for the resources they provided. R. Nichol thanks the Leverhulme
Trust for their support. K.D. de Vries is supported by
European Research Council under the European Unions
Horizon research and innovation program (grant agreement 763 No 805486). D. Besson, I. Kravchenko, and D. Seckel thank the NSF for support through the IceCube EPSCoR Initiative (Award ID 2019597). M.S. Muzio thanks the NSF for support through the MPS-Ascend Postdoctoral Fellowship under Award 2138121. A. Bishop thanks the Belgian American Education Foundation for their Graduate Fellowship support.
}

\end{document}